\documentclass[a4,10pt,twocolumn]{revtex4}    

\usepackage{graphicx}

\begin{document}

\thispagestyle{empty}

\title{\sffamily\Large\bfseries Optimization Aspects of Carcinogenesis}

\author{B. Brutovsky$^a$}
\author{D. Horvath$^{b,c}$}
\affiliation{           }
\affiliation{$^a$Department of Biophysics, P.~J.~Safarik University, Jesenna 5, 04154 Kosice, Slovakia}
\affiliation{$^b$Centre de Biophysique Mol\'eculaire, CNRS; Rue Charles Sadron, 45071 Orl\'eans, France}
\affiliation{$^c$Department of Physics, Faculty of Electrical Engineering and Informatics, 
Technical University, Letn\'a 9, 042 00 Ko\v{s}ice}

\begin{abstract}
{\it Any process in which competing solutions replicate with errors
and numbers of their copies depend on their respective fitnesses
is the evolutionary optimization process.
As during carcinogenesis mutated genomes replicate according
to their respective qualities, carcinogenesis obviously qualifies 
as the evolutionary optimization process and conforms
to common mathematical basis.
The optimization view accents statistical nature of carcinogenesis
proposing that during it the crucial role is actually played by 
the allocation of trials. Optimal allocation of trials requires reliable
schemas' fitnesses estimations which necessitate appropriate,
fitness landscape dependent, statistics of population.   
In the spirit of the applied conceptual framework, features which
are known to decrease efficiency of any evolutionary optimization procedure 
(or inhibit it completely) are anticipated  as "therapies" and reviewed.
Strict adherence to the evolutionary optimization framework leads
us to some counterintuitive implications which are, however, in agreement
with recent experimental findings, such as sometimes observed more
aggressive and malignant growth of therapy surviving cancer cells.
}
\end{abstract}

\maketitle

\section*{INTRODUCTION}

The term cancer refers to hundreds types of neoplasms which share specific
prototypical traits, summarized by Hanahan and Weinberg \cite{Hanahan2000},
collectively leading to malignant growth. During the past few decades molecular
biologists have produced much cancer-related data which has shown cancer
as an extremely stochastic, heterogeneous and complex disease \cite{Sjoblom2006}.
To analyze them, cancer research applies many concepts originally developed
in different branches of science, such as applied mathematics, nonlinear
dynamical systems, and statistical physics.
At present, evolutionary nature of carcinogenesis is accepted and
implications for cancer robustness (exemplified by resistance to therapy)
are often emphasized \cite{Komarova2003, Kitano2004b}.
Darwinian view to carcinogenesis implicitly puts genetic
(and epigenetic) changes into microenvironmental context \cite{Witz2006}.
Consequently, tumor microenvironment is viewed as an eventual target
for chemoprevention and cancer reversion \cite{Anderson2006, Ingber2008}.
On the other hand, anticancer research and therapy concentrate mainly 
on molecular data and tend to overlook its evolutionary nature.
 
Optimality model applied in experimental evolution \cite{Heineman2007}
describes the evolution as simple generalized trade-offs, presuming that genomes
adapt successfully and freely enough and, consequently, genetic details become
irrelevant. Mathematical approaches to carcinogenesis often apply
concepts of feedback and optimal control theory \cite{Chareyron2009}
instead of molecular or genetic data.
Komarova et al. \cite{Komarova2008} have solved the optimization problem
for cancerous growth and proposed optimal strategies. However,
as they state, the ideal (optimal) strategy may be not realistic due to 
many constraints in nature which escape modeling, but can make a strategy
impossible.

In the paper we concentrate on the abstract mechanisms
of attaining an optimal strategy instead of the strategy
itself. We view any process in which solutions replicate with errors and
numbers of their copies depend on their respective qualities
as an evolutionary optimization process.
As carcinogenesis conforms the above definition, we
identify it with an evolutionary optimization process
and apply concepts and results of the long lasting research
in the evolutionary optimization \cite{Holland1975}.
Keeping in mind an eventual therapeutic application, we focus on
those aspects of evolutionary optimization which decrease
or inhibit efficiency of the optimization process.
Strict adherence to the optimization framework 
has led us to counterintuitive implications.

\section*{EVOLUTIONARY OPTIMIZATION}

In the optimization theory, the quality of a solution is usually defined explicitly
in the form of a {\it fitness function} (also {\it fitness landscape} or {\it fitness}),
quantifying how well a candidate solution meets required criteria.
The ultimate aim of the optimization procedure is to find a solution
for which the fitness function receives optimum value. Large group
of optimization algorithms, called evolutionary algorithms (EA),
performs the task by mimicking biological evolution implementing
the genetic-like mechanisms, such as mutation, selection and
reproduction. Applying EA in various engineering optimization applications
has enabled to recognize those aspects of fitness landscapes which support
efficient evolutionary optimization and, at the same time, those which prevent it. 
Theoretical analysis of the most popular EA variant, the genetic algorithms (GA),
has been performed by J. H. Holland \cite{Holland1975}. 
In the simplest engineering applications, canonical GA (CGA)  applies:

\bigskip

{\noindent
1. initial population of random binary strings is generated
\begin{center}
                              1011101011 ...\\
                              0001100101 ...\\
                               . \\
                              1011110010 ...\\
\end{center}
\noindent
2. each of the bit string is projected and scaled to get the real parameter set ${\bf r}=(r_1,r_2,\cdots)$
$$
\underbrace{1001{\cdots}1010}_{r_1}\underbrace{0110{\cdots}0111}_{r_2}
\cdots 
$$
and its fitness function value  $\phi({\bf r})$   is determined

\noindent
3. child population of the bit strings is constructed from the parents population
applying the genetic operators - selection depending on the strings fitnesses, 
crossover and mutation; after it is complete it replaces the parent population

\noindent
4. until some convergence criterion applies go to the step 2.
}

\bigskip

Theoretical analysis of the process enabled to identify the driving force behind
the biological-like manipulations with binary strings representing parameters of the model.
It was recognized that the population-based optimization algorithm is driven by
the fitnesses of the correlations of bits in the binary strings (called "schemas").
The schema can be viewed as a bit pattern over the bit positions in the string.
If the bit alphabet \{0,1\} is assumed, the schema can be easily constructed over
the ternary alphabet \{0,1,*\}, where '*' matches both, 0 and 1, at the respective position:

\bigskip

{\noindent
Let's have 4 binary strings
\begin{center}
                            A   10100101\\
                            B   01011011\\
                            C   11100010\\
                            D   00010001
\end{center}
and two schemas, X and Y
\begin{center} 
                            X   *1***01*\\
                            Y   0*01***1
\end{center}
The schema X is contained in the strings B and C, the schema Y in the strings B and D.
It is usually said that strings B and C are instances of the schema  X and the strings
B and D are the instances of the schema Y.
}

\bigskip

Specificity and robustness of the schemas are quantified by the schema order, 
${\cal O}$, and defining length,  $\delta$.
The schema order is the number of fixed positions in the schema.
The defining length is the distance between the leftmost and the rightmost 
fixed positions.
To predict the number of instances of a schema in generation
$t+1$, Holland derived the schema theorem (ST) \cite{Holland1975}

\begin{equation}
\label{SchemaTheorem}
N^{t+1}\ge N^t\frac{\Phi^t}{\overline{\phi^t}}
\left[1-P_c\frac{\delta}{l-1}-{\cal O}P_m\right], 
\end{equation}

\noindent
where $N^t$ and $N^{t+1}$ are numbers of instances of the schema
in $t$ and $t+1$, respectively, $P_c$ is the probability of strings
crossover, $P_m$ is mutation rate, $l$ is the length of the binary string,
$\overline{\phi^t}$ is the average fitness in the population, and 
$\Phi^t$ is the schema fitness in $t$ defined as the average fitness of all
the instances of the schema in the population in $t$.
ST (\ref{SchemaTheorem}) states
that during the GA optimization the number of the above average
schemas increases on the account of less favorable schemas.
Moreover, it has been demonstrated, that GA allocates its trials among
alternative solutions during the search (known as $k$-arm bandit problem)
in optimum way as long as the schemas' fitnesses are correctly estimated
\cite{Holland1975}.

As in the paper we have identified carcinogenesis with the
evolutionary optimization process, any feature or mechanism
which decreases efficiency of the optimization process is interesting
from the point of view of its eventual therapeutic application.
Recognizing ST (\ref{SchemaTheorem}) as the principal mechanism
driving the evolutionary optimization, an explicitly
optimization-preventing therapy can be identified with
substituting into (\ref{SchemaTheorem}) wrong schemas'
fitnesses estimates. Below we list in the GA literature most
often presented reasons preventing reliable estimates
of the schemas' fitnesses.

{\noindent\it i) Too large sampling errors.}
The factors influencing the reliability of statistical sampling
are the number of evaluated candidate solutions
and their distribution in the search space (i.e. population heterogeneity).
They should cover as much of the search space (fitness landscape)
as possible so that the convergence to the optimum was as probable
as possible. The sampling errors can be reduced by
the appropriate choice of mutation rate. If mutation rate is too low,
optimization sticks in a suboptimal solution (known as premature
convergence). If mutation rate is too high, optimization
procedure turns into so-called blind search.

{\noindent\it ii) Dynamic fitness landscape.} As the evolutionary optimization
procedure converges towards optimum solution in a stationary fitness
landscape, heterogeneity of the population decreases.
The parts of the search space near the optimum become overpopulated,
and, at the same time, other parts only sporadically populated, or even empty. 
The role of the observed increase of population heterogeneity
in changed  environment is well interpretable using
the terms of evolutionary optimization, namely evolution algorithms
in dynamic environments. Therefore, mechanisms of heterogeneity
maintenance  have been developed in optimization theory and deeply studied
\cite{Morrison2004}. 
Efficient transition 
from the old optimum to optimum(a) in a new fitness landscape
requires i) detection of the fitness landscape change, and
ii) response to that change \cite{Morrison2004}.
For that, candidate solutions must be appropriately distributed
in the search space so that evolutionary algorithms could perform
representative statistical sampling to determine reliable schemas' fitnesses
estimates which are necessary for optimal allocation of trials during
optimization. If there are no (or too few) evaluations in the changed
part of the fitness landscape, the change goes undetected.

{\noindent\it iii) Deceptiveness of fitness landscape.}
To answer the question which fitness landscapes are
GA-hard, Bethke \cite{Bethke1980} expressed a fitness function
as a linear combination of Walsh monomials and showed
the relationship between the schema's fitness and Walsh coefficients.
Consequently, he applied the Walsh transform to characterize functions
as easy or hard for GA optimization. It has been understood that
the principal problem for GA optimization is the class of deceptive
fitness functions, in which lower order (lower number of defined bits)
schemas lead the search towards bad higher order schemas. Goldberg
showed the possibility of constructing high-order deceptive functions
using low-order Walsh coefficients in special cases \cite{Goldberg9}.

\section*{CARCINOGENESIS AS EVOLUTIONARY OPTIMIZATION PROCESS}

Exact convergence analysis of EA requires much better mathematical
definition of the relevant fitness landscape and more obvious parametrization
of a solution than one typically disposes with biological systems.
Regarding the above introduced schema formalism a few differences
between CGA and carcinogenesis should be mentioned. At first,
carcinogenesis is an asexual process, therefore constant $P_c$ in 
(\ref{SchemaTheorem}) equals zero.
The second difference is that no spatial relation between offsprings
and their parents is assumed in (\ref{SchemaTheorem}). The third
difference regards unknown parametrization - obviously higher
structures than nucleotides (or genes) are relevant. Nevertheless, 
neither of the differences puts in doubt importance of reliable
estimates of the schemas' fitnesses for optimal allocation of trials
during carcinogenesis. In addition, as often used in evolutionary optimization
practice, we use the term optimum solution in a sense of a winning solution,
i. e. the best solution obtained after reasonable (or affordable) long optimization,
instead of exact, mathematically proved, solution.

\begin{flushleft}
{\bf  Fitness landscape}
\end{flushleft}

The term represents central concept in biological evolution
as well as in optimization theory \cite{Beveridge1}. In biology,
the fitness is usually understood in a sense of "reproduction" fitness,
meaning that the more copies solution has the more fit it is (and vice versa),
and obtains factual meaning in specific environment and time scales.
During the genome's evolution selection acts at two different hierarchical
levels respective to the two units of replication: cells and organisms (multicellular bodies).
As a result, the genome is the trade-off between two processes:
i) maximization of the multicellular (organismic) reproduction fitness
(acting during millenia), and ii) maximization of cellular reproduction
fitness (acting during individual lifespan), respectively. The former process
presumes social cooperation of cells (such as limited replicative potential,
production of growth signals, sensitivity to antigrowth signals, cellular
senescence, apoptosis, etc.) and severe prohibition of the cells' selfishness,
the latter favors selfishness instead of cooperation \cite{Greaves2007}.
The trade-off is mediated by the initial genomic stability, evolved to postpone
short scale evolution in the respective environment
beyond reproduction period of the respective organism.

\begin{flushleft}
{\bf Heterogeneity}
\end{flushleft}

Extensive genomic studies by Sj\"oblom et al. \cite{Sjoblom2006} have
clearly demonstrated extreme heterogeneity in colorectal cancer tumors.
They have revealed that mutational patterns in samples of colorectal cancers
are unexpectedly individualistic, with none of the three most often
mutated genes (APC, p53, K-ras) mutated in all the samples
\cite{Beerenwinkel2007}. It has been shown that sets of mutated genes
in two samples of colorectal cancers overlap to only a small extent and
it is anticipated to be general feature of most solid tumors \cite{Wood2007}.
Similarly, resuming studies in breast and renal cancer,
Gatenby and Frieden \cite{Gatenby2004} concluded that probably
no prototypical cancer genotype exists and every tumor seems
to possess a unique set of mutations indicating that multiple
genetic pathways may lead to invasive cancer as would be expected in 
a stochastic non-linear dynamical system.
Clonal diversity in a subset of patients with early stage haematopoietic malignancy
has been demonstrated and it has been shown that such clones may arise
independently \cite{Beer2009}.
It has been also observed \cite{Taniguchi2008}, that time to disease
progression and overall survival after treatment were significantly
shorter in those patients with EGFR heterogeneity.
Maley et al. \cite{Maley2006} have demonstrated that clonal diversity
predicts progression to cancer and that accumulation of viable clonal
genetic variants is a greater risk for progressing to cancer than
homogenizing clonal expansion.
Mathematical model by Komarova et al. \cite{Komarova2008}
shows that tumors thrive when cancerous cells mutate to speed up malignant
transformation, and then stay that way by turning off the mutation rate.

Interpretation of heterogeneity is crucial for understanding of carcinogenesis.
It can be, in extreme cases, interpreted either as noise hiding a common pattern,
or redundancy (all the cases are causative as a whole, no common pattern exists).
If interpreted as a noise, the effort to filter it out by analyzing as many cancer
cases as possible to see the common mechanism is justified. If, however,
each sample is interpreted as a unique, nevertheless causative
set of genes, alternative approaches are needed. The above mentioned studies
at genetic level \cite{Sjoblom2006,Beerenwinkel2007,Greenman2007,Wood2007}
indicate that heterogeneity should be interpreted in the latter way.
They report that every tumor harbors a complex combinations
of low-frequency mutations thought to drive the cancer phenotypes \cite{Luo2009}.
Consequently, a strategy to study mechanisms of cancer by reducing
heterogeneity may be assumed to be a flawed approach \cite{Heng2009}.   

\begin{flushleft}
{\bf Optimization behind}
\end{flushleft}

Putting fitness landscape and heterogeneity into optimization context,
the wild-type genome represents optimum solution in the respective
past fitness landscape; its further optimization in unchanged fitness
landscape is, by definition, inhibited. After the fitness landscape
has changed, optimization of the genome becomes possible.
Regarding the structure of the fitness landscape, during the optimization
two fitness landscapes are sampled, each for the respective
unit of replication - organism or cell. As there are many cellular fitness
evaluations during the organism's lifetime, only cellular fitness landscape
may be sampled representatively enough to provide reliable schemas'
fitnesses (\ref{SchemaTheorem}) which result in optimal allocation 
of trials driving the short time evolution of the genome into an optimum
in the changed cellular fitness landscape. The organismic fitness landscape,
selecting for intercellular cooperation, does not apply during the lifetime
of the body and the optimization process is driven purely by cellular fitness
landscape for which the intercellular cooperation is not selectable trait.
From this point of view, any short-scale change of the fitness landscape
is not only mutagenic but also carcinogenic, as it selects for destroying
intercellular cooperation. 
Applying the quasispecies model \cite{Eigen1971}, Forster and Wilke
have demonstrated that competitive dynamics of finite populations
of as few as two strains, adapted to the long-term and short-term
environment changes, respectively, is quite complex \cite{Forster2005}.
 
Heterogeneity represents crucial aspect of carcinogenesis \cite{Heng2009}.
At the same time, in engineering applications, evolutionary optimization starts
with heterogeneous, typically randomly generated, initial population of candidate
solutions. In the case of stationary fitness landscapes, heterogeneity
decreases towards some minimum level as the optimization procedure
converges to the best solution (the analogy with a homogenizing clonal
expansion inflicts itself), despite keeping constant mutation rate.
On the other hand, evolutionary optimization in changing fitness landscapes
\cite{Morrison2004} shows importance of avoiding total homogenization.
In computer experiments where mutation rate is not exempted from optimization,
its increase (followed by the increase of heterogeneity) is observed
after the fitness landscape has changed. It has been reliably
demonstrated that rapid or extreme environmental change leads to the selection
for greater evolvability \cite{Earl2004}.
Similarly, selection of mechanisms for increased mutation rate in biological systems,
like RNA viruses, in unstable environments was reported \cite{Tenaillon1999}.
Donaldson-Matasci et al. \cite{Donaldson2008} have shown that
optimal amount of diversity depends on environmental uncertainty which
can lead to the evolution of either generalist or specialist strategy.

Cancer-susceptibility genes are classified as caretakers, gatekeepers
and landscapers. Mutations in caretakers leads to genomic instability,
mutated gatekeepers are responsible for increased cellular proliferation
and landscapers defects generate an abnormal stromal environment.
In general, the cancer-susceptible genes govern statistics
of the cell population, either directly (caretakers and gatekeepers),
or indirectly by maintaining fitness landscape (landscapers). 
Within the frame of evolutionary  theory it is understood that heterogeneity
confers cancer cells population with the ability to cope with environment
uncertainties. Optimization theory derives efficiency of an optimization
method from its ability to allocate appropriately future trials. 
The schema theorem (\ref{SchemaTheorem}) guarantees giving at least 
exponentially increasing number of trials to the observed best building blocks
\cite{Goldberg1989}. Implicitly, optimal allocation of trials between
alternative solutions requires as reliable schemas' fitnesses estimates
as possible. In addition, as evolving clones implicitly undergo competition,
the schemas' fitnesses must be determined as fast as possible.
For that, representative (regarding the respective fitness landscape) statistics
of the population must be at hand. The ability of the clone to evolve 
(or not) towards  representative statistics comes from specific defects
in cancer susceptibility genes.

Causality in evolutionary processes is actually provided by the feedback
from environment. The evolutionary process is a fitting procedure, which
is the method of solving (typically ill-posed) inverse problems
\cite{Sabatier1985}. Enormous genetic heterogeneity of cancers
indicates that most cancer occurrences are the unique solution of the
fitting problem. It implies that the fitting problem solved by cancer
is highly underdetermined, which results in the arbitrariness of a fit
(i. e. model) and it is consistent with the metaphoric conclusion
by Witz and Levy-Nissenbaum \cite{Witz2006}, who stated "...the extreme
complexity of the signaling cascades operating in the microenvironment
and the interactive cross-talk between these cascades, generates
the feeling that 'anything that can happen - it will".

\section*{IMPLICATIONS FOR THERAPY}

Traditional therapies are based on comparisons of cancerous and non-cancerous
cells, which, by definition, presumes existence of reliable enough (in an ideal case dichotomic)
splitting into two respective groups. Consequently, therapeutic actions are taken
to attack the tumor cells group (cancer cell-kill paradigm). It is implicitly
believed that therapeutic efficiency depends on how close
to dichotomic the splitting is.
For instance, the two main therapeutic treatments, chemotherapy and radiation,
exploit the enhanced sensitivity of cancer cells to DNA damage.
Novel targeted and gene therapies go even further - they are aimed to interfere
directly with the specific molecules or genes participating in carcinogenesis
(the 'magic bullet' concept).
The effort to find the criterion(a) enabling to approach to dichotomic
splitting as close as possible is omnipresent in cancer therapy.
Varshavsky \cite{Varshavsky2008} proposed the therapy which distinguishes
cancer and normal cells according to harboring (or not) homozygous
DNA deletions.
Skordalakes \cite{Skordalakes2009} points out that inappropriate activation
of a single enzyme, telomerase, is associated with the uncontrollable proliferation
of cells observed in as many as 90\% of all of human cancers and proposes
that the high-resolution structure of the enzyme will be the key to efficient
anti-cancer therapies.

However, putative existence of dichotomic splitting is in contradiction with
the evolutionary nature of carcinogenesis which, as any other evolutionary
process, crucially depends on the variability of traits observed at many levels
\cite{Fidler1978,Futreal2004,Sjoblom2006,Beerenwinkel2007}.
Extreme tumor cells heterogeneity gives cancer robustness, exemplified
by the resistance to therapy \cite{Komarova2003,Kitano2004b},
and it is the most tormenting problem in cancer research to which therapies
and experimental models must face \cite{Klein2002, Kitano2003}. 
Heng et al. \cite{Heng2009} emphasize the key role of heterogeneity by stating that
without heterogeneity, there would be no cancer. 

Below we present specific insights and implications for anti-cancer therapy
stemming from the above presented optimization view to carcinogenesis.
Some of them are intuitive and consistent with  established anti-cancer
therapies, some others are quite counterintuitive and,
hopefully, novel and put in question some current trends
in the development of anti-cancer therapies.
Within the frame of the above outlined identification of carcinogenesis
as the evolutionary optimization process, therapy is a purposeful
effort to decrease the efficiency of that optimization process or,
hopefully, inhibit it completely. For that purposes, we have listed above 
the three most frequent obstacles to the efficient evolutionary
optimization, stemming from validity of the schema theorem
(\ref{SchemaTheorem}). These are: too large sampling errors, dynamic
(or changing) fitness landscapes and deceptiveness of fitness landscape. 
In all the cases the estimation of the schemas' fitnesses is not reliable
(or systematically wrong) which prevents the optimization process 
to allocate its trials optimally. 

\bigskip

{\it i) Too large sampling errors.}
It is understood that heterogeneity plays a central role
in evolution and provides species (or clones) with the capacity
to cope with environmental uncertainty. On the other hand,
if it exceeds a certain threshold, deleterious effects outweigh
the above selection advantage.
The existence of the critical mutation rate in evolution beyond
which Darwinian selection does not operate has been predicted
by Eigen's theory of quasispecies \cite{Eigen1971}. 
Sole and Deisboeck \cite{Sole2004} applied the simple mathematical
model of quasispecies dynamics to quantify the upper limit
of affordable genetic instability (error threshold) in cancer cells
population, beyond which genetic information is lost.
Consistently with the fact that tumor cells have defective stability
pathways, Cahill et al. proposed that tumor cells could be target
for direct attack by instability drugs \cite{Cahill1999}.
However, from the point of view of the evolutionary optimization
theory, competitiveness of the clone depends on its capability
to allocate its further trials among emerging alternatives \cite{Holland1975}
which requires representative statistics of the cells population, not merely
specific genetic (in)stability. Therefore we speculate that
forced increase of sampling errors by instability drugs, abruptly
shifting population statistics away from the optimal in the respective fitness
landscape, would be compensated by selecting for change(s) in
other evolutionary attribute(s), such as reproduction rate, cellular mortality
rate, internal stability (the mechanism does not matter at this point), etc.

\bigskip

{\it ii) Dynamic fitness landscape.}
Changing the fitness landscape can be a double edged sword.
On the one hand, cancer cells reveal increased adaptivity enabling
them to respond to environmental changes to keep high (reproductive)
fitness. On the other hand, higher adaptivity of cancer
cells can be therapeutically exploited, as outlined by Maley et al.
\cite{Maley2004}. They proposed to select for the cells
sensitive to cytotoxins before applying cytotoxic therapy.

{\it iii) Deceptiveness of fitness landscape.}
Deceptive landscapes can be interpreted as
the landscapes in which correlations of traits systematically
lead away the search from the global optima.
To our knowledge, there is no therapeutic approach explicitly
exploiting deceptiveness of the fitness landscape.
We anticipate that combining biological intuition, the results
of mathematical analysis of deceptive fitness landscapes
\cite{Goldberg9} and
digitized evolution \cite{Wilke2001a} can bring novel
insights into the evolution of cancer phenotype. 

\begin{flushleft}
{\bf Is therapy a penalty function?}
\end{flushleft}

From the evolutionary optimization point of view therapy
is a purposeful change of the fitness landscape, namely decrease
of reproduction fitness in the relevant area of the search (sequence)
space at reasonable time scales. All the well established traditional
therapies (surgery, radiotherapy and chemotherapy) make an effort
to remove all the cancer cells, or, at least, as many of them as possible.
Evolutionary optimization theory implies that ultimate therapeutic success
depends not only on how many cancer cells survived the therapy, but also
on the distribution of the cells in the search space, i. e. statistics
of the remaining population. If the population statistics is sufficient
for the efficient optimization, the regrowth appears.
Below we present eventual counterintuitive consequence of therapy
resulting from the optimization facet of carcinogenesis.

It has been reported that therapy-surviving tumor cells are frequently
more malignant and aggressive than the initial tumor population 
\cite{Schneider2004}.
Inhibition of angiogenesis has been envisioned as promising
anticancer therapeutic strategy for a long time \cite{Folkman1971}.   
Since then, modes of resistance to antiangiogenic therapy, such as
evasive and intrinsic resistance, has been reported \cite{Bergers2008}.
It has been found by Paez-Ribes at al. \cite{Paez-Ribes2009} that targeting
the vascular endothelial growth factor (VEGF) induces (apart from anti-tumor effects
to primary tumor) higher invasiveness and, in some cases, increased lymphatic
and distant metastasis. Ebos et al. \cite{Ebos2009} 
have found that the VEGFR/PDGFR kinase inhibitor can accelerate
metastatic tumor growth and decrease overall survival
in mice receiving short-term therapy.
Similarly, it has been reported that the resistance to some synergistic
drug combinations evolves faster than the resistance to individual drugs
\cite{Hegreness2008}.
In their review Kim and Tannock \cite{Kim2005} report that repopulation
of cancer cells after radiotherapy as well as chemotherapy is often
accelerated in comparison to untreated cases.
The mechanism of this acceleration has not yet been understood. 

\begin{figure}[!h]
\includegraphics[scale=0.8]{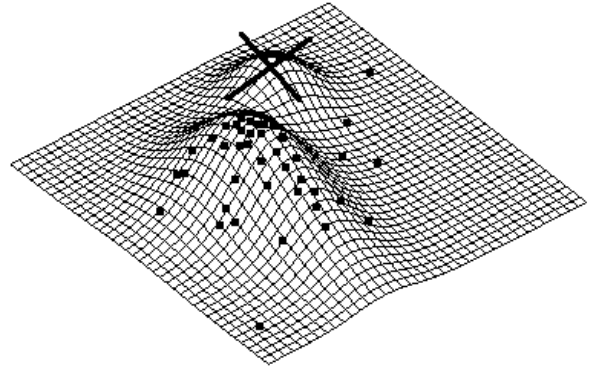}
\caption[]{\noindent\small 2-dimensional fitness landscape with one "hill" (crossed)
prohibited by a penalty function. Dots show sampled points. Without penalty, both hills would be
sampled.}
\label{penalty}
\end{figure}

In the spirit of our work we attribute the above increase of invasiveness
and acceleration of the evolution of resistance during repopulation
to the optimization facet of carcinogenesis.
In engineering applications of evolutionary optimization
one often applies {\it ad hoc} penalty function to disadvantage
some part(s) of fitness landscape to accelerate convergence
of the process into the optimum in desirable parts (Figure~\ref{penalty}).
The simplification of fitness landscape enables to perform more representative
schema sampling of more promising parts at the same price obtaining
more reliable schemas' fitnesses evaluations resulting,
accordingly to (\ref{SchemaTheorem}), in closer-to-optimum
allocation of the trials among alternative solutions.
If cancer, metaphorically said, solves the optimization problem,
the same mechanism applies.
We hypothesize, that if therapy does not remove decisive
portion of cancer cells (hopefully all), it may, eventually, result
in unwanted simplification of fitness landscape for therapy-resistant
clone(s). We emphasize, that this hypothesis is aimed purely to
interpret sometimes reported cases when accelerated progression of
therapy-resistant tumors was observed and it does not propose
any alternative to well established therapies.

\section*{DISCUSSION}

Recent experimental evidence shows that heterogeneity, stochasticity
and dynamics play in carcinogenesis much more important role than
envisioned a few decades ago. This new picture requires corresponding
conceptual framework. Here presented evolutionary optimization view
to carcinogenesis implicitly includes connection between statistics
of cells population and statistics of fitness landscape
\cite{Forster2005,Nilsson2002,Tanaka2003,Travis2002,Clune2008,Brunet2008}
and applies results of long-standing research in the stochastic evolutionary
optimization algorithms, especially in dynamic fitness landscapes
\cite{Morrison2004,Grefenstette1999}.
Here we have put some of the observed cancer features, such
as increased heterogeneity, clonal expansion, consequences of changing
the fitness landscape and accelerated evolution of resistance to
chemotherapy into optimization scenario. 

Carcinogenesis is, unquestionably, a physical process. At the same time,
it can be formally viewed, as all the evolutionary processes, as the  optimization
procedure. Straightforward approaches study carcinogenesis and develop
anticancer strategies analyzing biochemical or genetic details.
In the paper we have speculated that it may be not relevant {\it per se}.
Instead, we have proposed that cancer relates primarily to
the cells population statistics and all the therapies lead
(more or less intentionally or explicitly) to its modification.
Traditional therapies rely on comparison between cancerous and
non-cancerous cells which may be motivated by the long lasting
effort to reduce cancer cells population by some straightforward action.
Evolutionary view suggests that carcinogenesis could be inhibited
by a purposeful modification of evolutionary attributes, such as mutation
rate, effective population size or generation time of the self-­renewing
cells \cite{Pepper2009}.
Nevertheless, except for trivial cases, evolutionary theory does not give
instructive  enough answer how should be the evolutionary attributes
changed.
Here presented optimization view to carcinogenesis
proposes that the crucial mechanism of cancer progression is, as in any
other evolutionary optimization process, optimal (or, more realistically,
better than by other clones) allocation of trials, based on more
representative population statistics enabling more reliable estimations
of schemas' fitnesses (\ref{SchemaTheorem}).
Efficiency of the schema sampling depends on the number of sampled
points and their distribution in the fitness landscape, as well as the
cell's fitness estimation time. These attributes adapt to statistical features  
of the fitness landscape by selecting respective mutations in genes
(a posteriori denoted as cancer-susceptible genes).
As, at the same time, efficiency of the sampling determines
the cancer's perspective, we conclude that the therapeutic outcome
could be influenced by manipulation with statistical properties of the
fitness landscape, such as roughness or dynamics, in a purposeful
cancer-inhibiting way.
The above statistical view may be relevant especially for advanced
malignancies, where high heterogeneity of the cancer cells population
enables them to adapt successfully to therapeutically-changed
environment.
Classifying carcinogenesis as the evolutionary optimization process
does not contradict to often presented view of cancer as the result
of accumulating specific mutations in the only transformed cell.
It emphasizes, however, importance to combine molecular data
with statistical view which may play crucial role before and
during carcinogenesis. The principal question
remains if the novel conceptual framework can be exploited
to trigger novel, explicitly anti-optimization based, therapeutic
approach.

\section*{ACKNOWLEDGEMENTS}

The authors acknowledge financial support
from VEGA, Slovak Republic (Grants 1/4021/07, 1/0300/09)
and APVV (grant LPP-003006).
D.H. acknowledges financial support by the {\sc LeStudium}
fellowship of the Region Centre and the
Centre National de la Recherche Scientifique.


\end{document}